\documentclass[aip,jap,reprint]{revtex4-1}
\usepackage[ansinew]{inputenc} 
\usepackage{graphicx}
\usepackage{amsmath}
\usepackage{SIunits}
\usepackage{xcolor}
\usepackage{nicefrac}
\usepackage{hyperref}
\begin{document}
\title{Spatially Resolved Determination of Thermal Conductivity by Raman Spectroscopy
\newline
\vspace{1cm}
\textit{accepted in Semicond. Sci. Technol.}
}
\author{B. Stoib}
\email[Electronic mail: ]{benedikt.stoib@wsi.tum.de}
\author{S. Filser}
\affiliation{Walter Schottky Institut and Physik-Department, Technische Universit\"at M\"unchen, Am Coulombwall~4, 85748 Garching, Germany}

\author{J. St\"otzel}
\affiliation{Nanostrukturtechnik, Faculty of Engineering, Universität Duisburg-Essen, Bismarckstr.~81, 47057 Duisburg, Germany}

\author{A. Greppmair}
\affiliation{Walter Schottky Institut and Physik-Department, Technische Universit\"at M\"unchen, Am Coulombwall~4, 85748 Garching, Germany}

\author{N. Petermann}
\affiliation{Institut f\"ur Verbrennung und Gasdynamik, Universität Duisburg-Essen, Carl-Benz-Str.~199, 47057 Duisburg, Germany}

\author{H. Wiggers}
\affiliation{Institut f\"ur Verbrennung und Gasdynamik, Universität Duisburg-Essen, Carl-Benz-Str.~199, 47057 Duisburg, Germany}
\affiliation{Center for NanoIntegration Duisburg-Essen (CENIDE), Universität Duisburg-Essen, Bismarckstr.~81, 47057 Duisburg, Germany}

\author{G. Schierning}
\affiliation{Nanostrukturtechnik, Faculty of Engineering, Universität Duisburg-Essen, Bismarckstr.~81, 47057 Duisburg, Germany}
\affiliation{Center for NanoIntegration Duisburg-Essen (CENIDE), Universität Duisburg-Essen, Bismarckstr.~81, 47057 Duisburg, Germany}

\author{M. Stutzmann}
\author{M. S. Brandt}
\affiliation{Walter Schottky Institut and Physik-Department, Technische Universit\"at M\"unchen, Am Coulombwall~4, 85748 Garching, Germany}

\date{\today}
\begin{abstract}
We review the Raman shift method as a non-destructive optical tool to investigate the thermal conductivity and demonstrate the possibility to map this quantity with a micrometer resolution by studying thin film and bulk materials for thermoelectric applications. In this method, a focused laser beam both thermally excites a sample and undergoes Raman scattering at the excitation spot. The temperature dependence of the phonon energies measured is used as a local thermometer. We discuss that the temperature measured is an effective one and describe how the thermal conductivity is deduced from single temperature measurements to full temperature maps, with the help of analytical or numerical treatments of heat diffusion. We validate the method and its analysis on 3- and 2-dimensional single crystalline samples before applying it to more complex Si-based materials. A suspended thin mesoporous film of phosphorus-doped laser-sintered $\rm{Si}_{\rm{78}}\rm{Ge}_{\rm{22}}$ nanoparticles is investigated to extract the in-plane thermal conductivity from the effective temperatures, measured as a function of the distance to the heat sink. Using an iterative multigrid Gauss-Seidel algorithm the experimental data can be modelled yielding a thermal conductivity of \unit{0.1}{\watt\per\meter\usk\kelvin} after normalizing by the porosity. As a second application we map the surface of a phosphorus-doped 3-dimensional bulk-nanocrystalline Si sample which exhibits anisotropic and oxygen-rich precipitates. Thermal conductivities as low as \unit{11}{\watt\per\meter\usk\kelvin} are found in the regions of the precipitates, significantly lower than the \unit{17}{\watt\per\meter\usk\kelvin} in the surrounding matrix. The present work serves as a basis to more routinely use the Raman shift method as a versatile tool for thermal conductivity investigations, both for samples with high and low thermal conductivity and in a variety of geometries.
\end{abstract}
\maketitle
%
%
\section{Introduction}\label{sec:Introduction}
In many fields of materials research and development, heat management becomes increasingly important. Both, exceptionally high or low thermal conductivities may be required for optimum device functionality. For example, in microelectronics it is required to efficiently cool integrated circuits to avoid diffusion or electromigration, thus making high thermal conductances on a sub-micrometer scale necessary.\cite{Tong2011,Moore2014,Balandin2009,Balandin2011} Graphene or isotopically purified crystals have been proposed as useful high thermal conductivity materials for such applications.\cite{Morelli2002,Balandin2011} On the other hand, thermoelectric devices or sensors based on micro-calorimetry benefit from materials with a low thermal conductivity, capable of sustaining temperature differences.\cite{Niklaus2007,Snyder2008,Kanatzidis2010,Nielsch2011,Schierning2014} To this end, material inhomogeneities on the micro- and nanometer scale can help to efficiently block heat transport by phonons due to wavelength selective scattering.\cite{Hochbaum2008,Tang2010,Biswas2012} 

Along with the increasing importance of thermal management, advanced techniques are being developed to experimentally measure thermal conductivities. Standard methods used today include the laser flash method for samples of rather large dimensions and well defined thickness,\cite{Cape1963} the $3\omega$~method for flat thin films with a good thermal junction to the underlying substrate,\cite{Borca-Tasciuc2001} micro-electromechanical measurement platforms, e.g., for individual samples of nanowires,\cite{Voelklein2010} or time or frequency domain thermoreflectance measurements for samples with well defined specular and temperature dependent reflectivity.\cite{Cahill2004} Hardly any of these techniques are free of challenges, such as limited throughput, unknown heat capacity, rough sample surface, highly diffusive reflection, high electrical conductivity, poorly defined sample thickness or spurious thermal conductance by contacts, substrates or the ambient.\cite{Tritt2004}

Especially in the regime of materials with low thermal conductivity, micro- and nano\-struc\-tures offer the possibility to reduce thermal transport.\cite{Yang2012} Thus, obtaining information on the local thermal conductivity is key to understanding and optimising materials properties, but is also rather demanding. Force microscopy methods are suitable to extract local differences of the thermal conductivities, but the quantification remains difficult.\cite{Nonnenmacher1992,Fiege1999,Meckenstock2008,Majumdar1999,Gomes2007,Zhang2010} Local measurements of the thermal conductivity have also been reported using thermoreflectance methods.\cite{Huxtable2004,Zhao2012,Zheng2007,Wei2013}

Another optical method which is capable of measuring the thermal conductivity of materials is the Raman shift method, which is also called Raman thermography, micro Raman method or optothermal Raman measurement technique.\cite{Cai2010,Lee2011,Perichon2000,Balandin2008,Huang2009,Soini2010,Li2009,Doerk2010,Balandin2011} Using a strongly focused laser beam, this technique potentially offers a spatial resolution on the micrometer scale. Although this technique based on Raman spectroscopy was applied already to porous low-thermally conducting materials quite a few years ago,\cite{Perichon2000} it only became popular after the work of Balandin and co-workers for measuring the thermal conductivity of suspended graphene.\cite{Balandin2008,Ghosh2009,Teweldebrhan2010,Ghosh2010,Balandin2011,Chen2012,Nika2012,Yan2013} It has now been used by many groups and extended to other materials, such as carbon nanotubes, Si, SiGe, Ge or GaAs.\cite{Cai2010,Lee2011,Liu2013,Stoib2014,Soini2010,Chavez-Angel2014,Liu2011} The method uses the fact that the energy of Raman active phonon modes usually is dependent on temperature. If this dependence is known, the Raman spectrum obtained contains quantitative information on how strongly the sample was heated by the Raman excitation laser during the measurement, which, for a known excitation power, contains explicit information on the thermal conductance of the structure or device investigated. Together with sufficient knowledge about the sample geometry and the path of heat flow in the sample, it is possible to obtain the thermal conductivity $\kappa$, the material specific intensive quantity of interest.

The present work summarizes the theoretical and analytical basis of the Raman shift method and applies it to some complex structures and sample morphologies. In section~\ref{sec:RamanShift} we start by discussing in detail how a temperature can be measured by Raman spectroscopy and how it can be  simulated numerically. We present the principles of the Raman shift method by means of a one-dimensional model and introduce two-dimensional Raman shift mapping. In section~\ref{sec:ModelSystems} we obtain $\kappa$ of a homogeneous bulk material and of a thin suspended membrane. These examples are a preparation for section~\ref{sec:ApplicationofRSM}, where we apply the methods presented to structurally more complex systems, such as inhomogeneous bulk-nanocrystalline Si and a thin suspended mesoporous film made from SiGe nanocrystals, before closing with some concluding remarks.
%
%
%
%
%
\section{The Principle of the Raman Shift Method}\label{sec:RamanShift}\label{sec:PrincipleRSM}
We start our introduction into the fundamentals of the Raman shift method by a discussion of how a temperature can be determined with Raman spectroscopy. Then, we use an  illustrative one-dimensional system to determine the thermal conductivity from such Raman temperature measurements, that serves as a model for our future studies of more complex sample structures.
%
%
%
%
%
\subsection{Measurement of an Effective Raman Temperature}\label{sec:HowtomeasureT}
\begin{figure}[tb]\centering\includegraphics{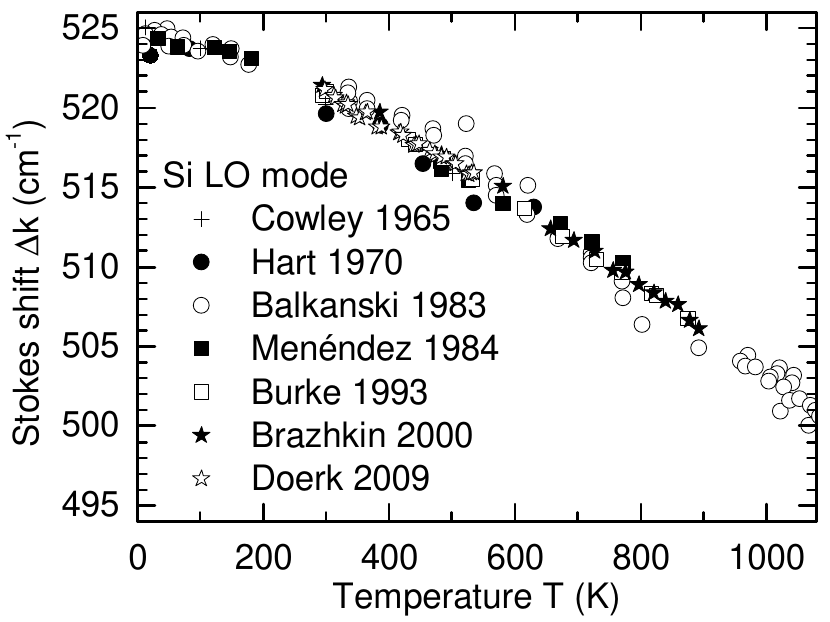}\caption{Temperature dependence of the Si longitudinal optical (LO) mode as reported in references \citenum{Cowley1965,Hart1970,Balkanski1983,Menendez1984,Burke1993,Brazhkin2000,Doerk2009}. Reproduced with permission from \href{http://dx.doi.org/10.1063/1.4873539}{Appl. Phys. Lett. \textbf{104}, 161907}. Copyright 2014, AIP Publishing LLC.}\label{fig:SiLO}\end{figure}

In harmonic approximation the energy of atomic vibrations in a solid is determined by the mass of the atoms and by the force constants between the masses. The anharmonicity of the potential leads to a change in the force constants with temperature and usually a crystal \textit{softens} with increasing temperature. In Raman scattering light interacts with these vibrations. Hence, the energy shift $\Delta k$ of Stokes and anti-Stokes scattered light also usually decreases with an increasing temperature of the sample studied.\cite{Cardona1983} In fact, the Stokes shift follows a distinct material specific dependence on temperature and can thus be used as a non-contact thermometer. The Stokes/anti-Stokes intensity ratio yields similar temperature information,\cite{Compaan1984a} but is, however, often more difficult to measure.\cite{Herman2011} As a typical example, the dependence of the Stokes shift $\Delta k$ of crystalline Si is shown in figure~\ref{fig:SiLO} for the longitudinal optical phonon mode.\cite{Cowley1965,Hart1970,Balkanski1983,Menendez1984,Burke1993,Brazhkin2000,Doerk2009} The choice of the phonon mode to be evaluated for temperature measurements depends mostly on the signal-to-noise ratio, but may also be influenced by the substrates available when investigating, e.g., thin films, since the Raman signal from the substrate should not interfere. Dependencies similar to figure~\ref{fig:SiLO} are observed in other solids as well,\cite{Menendez1984,Liu1999,Cui1998,Li2009a,Sahoo2013} making the Raman shift method applicable to a large variety of materials systems.

In the great majority of Raman spectroscopy experiments, the temperature distribution $T(\vec{r})$ is not homogeneous in the sample region where the laser light is Raman scattered. This means that the Raman spectrum collected will contain contributions of hotter (e.g., in the beam centre) and colder (edge of the laser beam) regions of the sample, caused by the inhomogeneous excitation via, e.g., a gaussian laser beam, \textit{and} the thermal conductance of the device studied. Thus, care must be taken when deducing a temperature from a Raman spectrum and the spectrum collected should be interpreted as a weighted average.\cite{Herman2011,Liu2011} We will call the temperature deduced from the Stokes shift $\Delta k$ measured an effective Raman temperature $T_{\rm{Raman}}$, to distinguish it from the local temperature $T(\vec{r})$ of the sample. In a very simple approach we assume that every location $\vec{r}$ on the sample contributes to $T_{\rm{Raman}}$ by its local temperature $T(\vec{r})$ weighted by the local excitation power density $H(\vec{r})$. We add up all those contributions in the sample volume and normalize it by the total absorbed laser power $P$ to obtain
\begin{equation}T_{\rm{Raman}}=\frac{1}{P}\int{H\left(\vec{r}\right) T(\vec{r})} c(T(\vec{r})) g(\vec{r})\text{d}\vec{r},\label{eq:weightingcomplicated}\end{equation}
where $c(T(\vec{r}))$ is the (in principle temperature dependent) Raman scattering cross section and $g(\vec{r})$ is a function that accounts for the effect that Raman scattering of weakly absorbed light takes place deep in the sample and that such scattered light is less efficiently collected by the objective. In all following calculations and experiments we will assume $c(T)$ to be constant. We further assume full surface near absorption, so that $g(\vec{r})=1$. Then, equation~(\ref{eq:weightingcomplicated}) simplifies to
\begin{equation}T_{\rm{Raman}}=\frac{1}{P}\int{H(\vec{r}) T(\vec{r})} \text{d}S,\label{eq:weighting}\end{equation}
where $\text{d}S$ is a surface element on the sample. This approach to determine $T_{\rm{Raman}}$ does also neither include the line-shape of the Raman signal nor its temperature dependence,\cite{Liu2011} but nevertheless improves the understanding of the Raman shift method in comparison to most analyses in literature and corrects the effects of different temperatures beneath the laser beam to first order.

\begin{figure}[tb]\centering \includegraphics{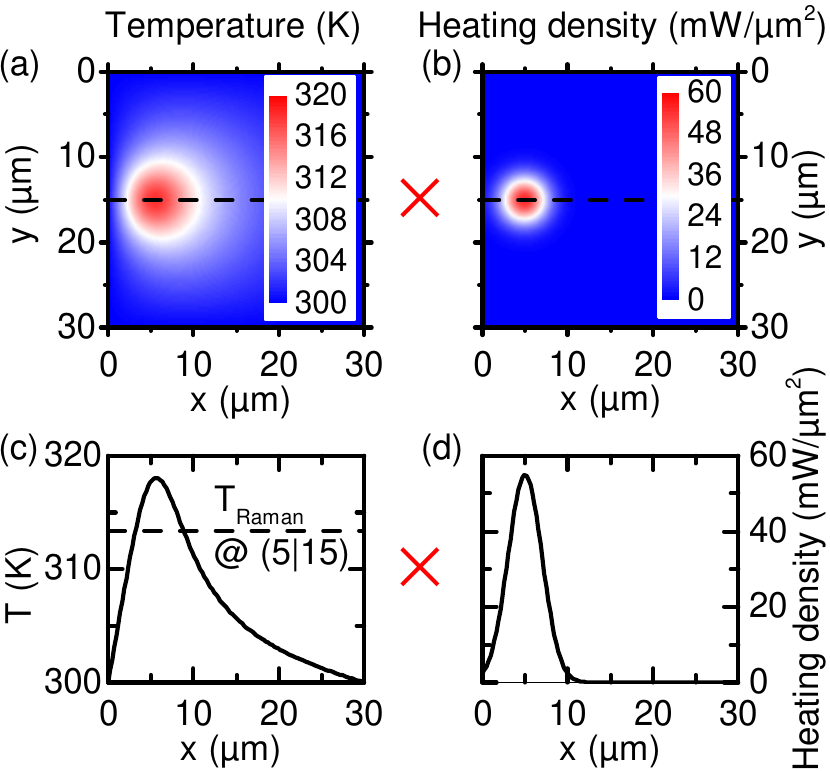}\caption{Weighting of the local temperature distribution with the excitation heating power distribution to obtain the Raman temperature $T_{\rm{Raman}}$. Panel~(a) shows a colour coded plot of a simulated temperature distribution on a $30 \times \unit{30}{\micro\meter^2}$ grid. A laser beam with a total absorbed power of \unit{100}{\milli\watt} and a standard deviation of \unit{2}{\micro\meter} excites a hypothetical 2-dimensional film with $\kappa=\unit{400}{\watt\per\meter\usk\kelvin}$ at $(x|y)$=(\unit{5}{\micro\meter}$|$\unit{15}{\micro\meter}). At the border of the film, the heat sink forces the temperature to \unit{300}{\kelvin}. Panel~(b) shows a colour coded plot of the gaussian heating power density. The temperature distribution in panel~(a) is the result of the excitation in panel~(b). Panel~(c) and (d) show cross sections of the colour plots at $y=\unit{15}{\micro\meter}$. Notably, the temperature is not constant in the area of excitation. The effective Raman temperature is a weighted average of the temperature distribution on the surface and the excitation laser power density and is shown in panel~(c) by the dashed line.}	\label{fig:Faltung}\end{figure}
Figure~\ref{fig:Faltung} shows an illustrative example of a hypothetical 2-dimensional square sample with $\kappa=\unit{400}{\watt\per\meter\usk\kelvin}$, which is heated by a gaussian laser beam at ($x|y$)=(\unit{5}{\micro\meter}$|$\unit{15}{\micro\meter}), having a standard deviation of $w=\unit{2}{\micro\meter}$. Panel~(a) shows a simulation of the temperature distribution $T(\vec{r})$ which is established in equilibrium on the square sample when exciting with the heating power density $H(\vec{r})$ shown in panel~(b). Panel~(c) and (d) are sections along the dashed lines in panel~(a) and (b), respectively. In the case shown, the Raman temperature according to equation~(\ref{eq:weighting}) at $(x|y)=(\unit{5}{\micro\meter}|\unit{15}{\micro\meter})$ is $T_{\rm{Raman}}=\unit{313}{\kelvin}$, which is significantly lower than the maximum temperature of \unit{318}{\kelvin}. 

The temperature distribution in figure~\ref{fig:Faltung}(a) obeys the stationary heat diffusion equation\cite{Carslaw1986}
\begin{equation}- H(\vec{r})=\kappa(\vec{r}) \Delta T(\vec{r})  +  \vec{\nabla}T(\vec{r}) \cdot \vec{\nabla}\kappa(\vec{r}).\label{eq:HDE}\end{equation}
It can be used because typically the minimum acquisition time of a Raman spectrum is of the time scale of a second, so that for small scale samples the measurement conditions are close to equilibrium. In equation~(\ref{eq:HDE}), a locally varying thermal conductivity $\kappa(\vec{r})$, e.g., due to a temperature dependent thermal conductivity, is considered.

For most sample geometries the temperature distribution for a given excitation cannot be calculated analytically. Whenever this is not possible, we use a numerical approach, where the field of interest is discretized in a rectangular grid and the discretized stationary heat diffusion equation is solved on every grid point. As an example we discuss a two-dimensional quadratic grid of dimension $a$, divided into $n$ grid points in each direction, so that one pixel has a width of $h =\frac{a}{n}$. The spatial coordinates $x$ and $y$ can be expressed by two indices $i$ and $j$
\begin{equation}
  (x,y) \rightarrow (i \times h,j \times h).
\end{equation}
Derivatives in equation~(\ref{eq:HDE}) are expressed in terms of discrete differences, e.g. 
\begin{equation}
 \frac{\partial^2 T}{\partial x^2} \rightarrow \frac{T_{i+1,j}+T_{i-1,j}-2T_{i,j}}{h^2} .
\end{equation}
The boundary conditions of a constant temperature $T_{\rm{sink}}$ outside of the simulation area and the continuity of heat flow, so that the heat introduced by $H(\vec{r})$ and the heat flowing into the heat sink are the same, are included. A thermal resistance $R_{\rm{th}}$ to the heat sink can also be considered. The problem to be solved can then be written as 
\begin{equation}
\underline{A} \cdot T_{i,j}=H_{i,j},\label{eq:MatrixFormDiscreteProblem}
\end{equation}
which is a linear set of equations with the matrix $\underline{A}$ containing all thermal conductivities and contact resistances.

Instead of directly solving equation~(\ref{eq:MatrixFormDiscreteProblem}), computation speed is enhanced by implementing a solver, based on an iterative Gauss-Seidel algorithm, where the computation effort only scales almost proportionally to the number of grid points.\cite{Briggs2000} In this algorithm the differential equation is not solved for all grid points simultaneously, but for each grid point in successive cycles, so that the discretized heat equation on each point is solved for $T_{i,j}$ with the values of neighboring points inserted from the previous cycle. This is repeated until a desired accuracy is achieved. Since spatially slowly varying temperature distributions converge only weakly, we use a multigrid algorithm on several grid sizes, first approximating the global temperature distribution on a coarse grid, and then refining this grid by factors of 2 and interpolating the temperature distribution stepwise.\cite{Briggs2000} Between all steps, Gauss-Seidel iterations are performed. The use of different grid sizes drastically speeds up the convergence of the method. 
%
%
%
%
%
\subsection{Determination of the Thermal Conductivity}\label{sec:Howtoobtainkappa}
\begin{figure}[tb]\centering \includegraphics{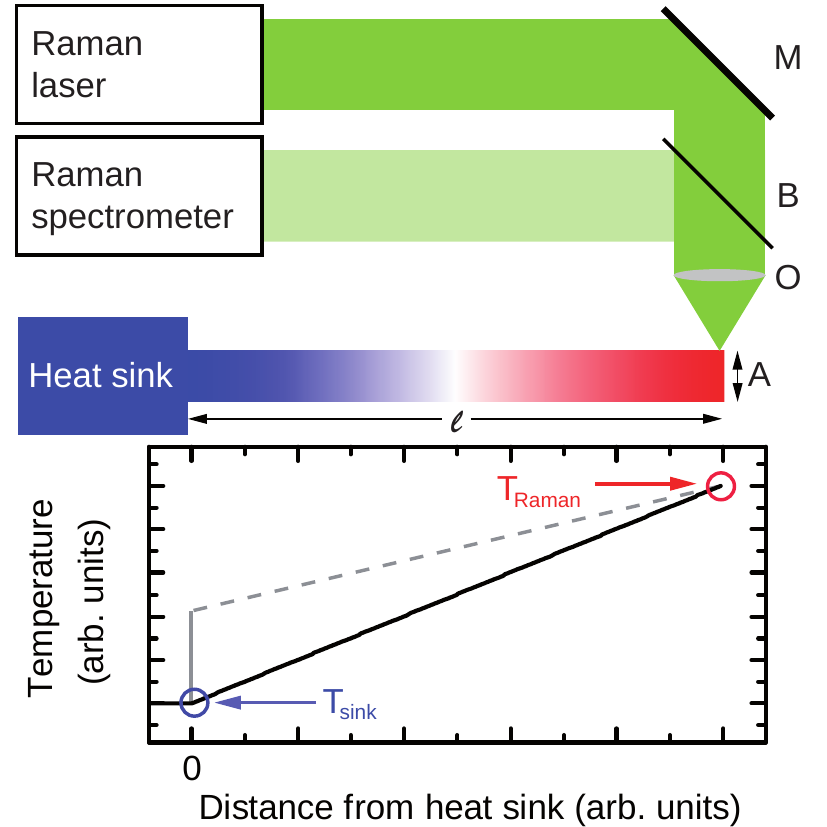}	\caption{Measuring the thermal conductivity of a bar-shaped material by the Raman shift method. The Raman laser acts both as the heating source and, together with the Raman spectrometer, as the thermometer. The beam of the laser is directed by mirrors (M) to the microscope objective (O), which focuses the light on the sample of length $l$ and cross section $A$. Raman scattered light is directed via a beam splitter (B) to the Raman spectrometer and $T_{\rm{Raman}}$ is measured. For vanishing contact resistance to the heat sink, the temperature distribution drawn as the black solid line is established in equilibrium. The grey dashed line considers a finite contact resistance to the heat sink and a lower thermal conductivity, so that the same Raman temperature would be measured at the end of the bar.}\label{fig:SchemaDidaktik}\end{figure}

By using the example of an effective 1-dimensional bar which is attached to a heat sink at one end, we will now discuss how the thermal conductivity of a material under test can be obtained based in the measurement of $T_{\rm{Raman}}$ introduced above. Figure~\ref{fig:SchemaDidaktik} schematically shows the focused Raman laser hitting the bar at its end and acting as the heat source. The heat generated at the right end will propagate through the bar to the heat sink on the left. For simplicity, let us assume that $\kappa$ in the bar is neither dependent on temperature nor position. Then, outside the laser beam where $H(x)=0$, equation~(\ref{eq:HDE}) can be written as
\begin{equation}0=\kappa \frac{\partial^2 T}{\partial x^2}. \label{eq:oneDimHDE}\end{equation}
Thus, the temperature decreases linearly from the excitation spot to the heat sink, as shown by the solid line in figure~\ref{fig:SchemaDidaktik}. 

To quantitatively obtain the thermal conductivity from equation~(\ref{eq:oneDimHDE}) and from the experimental value of $T_{\rm{Raman}}$, appropriate boundary conditions have to be set. As already pointed out, the continuity equation requires that the total heat generated at the bar's right end has to propagate to the heat sink. Neglecting the extension of the laser beam and a thermal contact resistance between the bar and the heat sink, the temperature of the bar at its left end is equal to the temperature of the heat sink $T_{\rm{sink}}$, so that 
\begin{equation}P=\frac{A}{l}\kappa \left(T_{\rm{Raman}}-T_{\rm{sink}}\right),\label{eq:oneDimCont}\end{equation}
where $P$ is the absorbed power, $A$ is the cross section and $l$ the length of the bar. This directly leads to
\begin{equation}\kappa=\frac{l}{A}\frac{P}{\left(T_{\rm{Raman}}-T_{\rm{sink}}\right)}.\label{eq:kappaOneDim}\end{equation}
In the example discussed so far $\kappa$ can be determined by only a single temperature measurement at the right end of the bar. If a thermal contact resistance has to be considered, at least a second temperature measurement needs to be performed at a different spot along the bar and equation~(\ref{eq:kappaOneDim}) has to be suitably changed. A possible temperature distribution along the bar for the case of a finite contact resistance is shown as a grey dashed line in figure~\ref{fig:SchemaDidaktik}. Eventually, performing many measurements along the bar, together with modelling the heat transport for the given sample geometry, significantly improves the accuracy of the method. 

\begin{figure}[tb]\centering \includegraphics{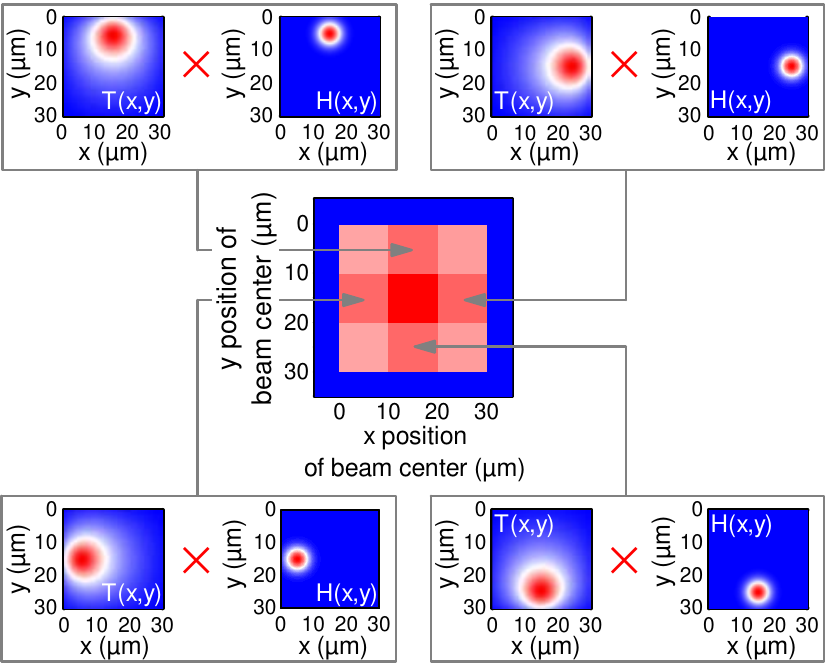}\caption{Simulation of a Raman temperature map. The laser beam is scanned across a sample and on every position, the effective Raman temperature is obtained by weighting the equilibrium temperature distribution $T(\vec{r})$ (left) with the local heating power density $H(\vec{r})$ of the excitation laser (right).}	\label{fig:MappingDidaktik}\end{figure}

In general, such spatially resolved temperature measurements can be performed in two dimensions scanning the whole sample surface, leading to what we will call a Raman temperature map. Figure~\ref{fig:MappingDidaktik} illustrates the generation of such a map by simulation. For each laser position on the sample surface the local temperature distribution $T(\vec{r})$ has to be calculated and weighted with $H(\vec{r})$ to obtain $T_{\rm{Raman}}$ at this spot. Experimentally, at each position a Raman spectrum is collected, and, using a relation such as the one shown in figure~\ref{fig:SiLO}, the corresponding effective temperature is deduced. By mapping the sample, enough information is collected to model both the thermal conductivity and a thermal contact resistance to the heat sink. Because the excitation as well as the temperature measurement are performed with a single laser beam, it is important to note, that such a Raman temperature map is not a temperature distribution, which via Raman scattering could be obtained only by using two lasers.\cite{Reparaz2014} There, the temperature distribution excited by a strong laser would be probed using a rather weak second laser, keeping the additional heating by the second laser to a minimum. 
%
%
%
%
%
\section{Model Systems}\label{sec:ModelSystems}
Before applying the Raman shift method to two material systems relevant for thermoelectrics, we first validate the method using bulk and thin film samples of single-crystalline Si and Ge.
%
%
%
%
%
%
\subsection{Heat Conduction Into a Semi-Infinite Half Space}\label{sec:3D}
The first model system is a homogeneous and semi-infinite bulk material, filling the half-space $z>0$. We want to analyse this system analytically and use cylindrical coordinates $r$, $\phi$ and $z$ to describe it. The Raman laser beam exhibits a radially gaussian shaped excitation power density
\begin{equation}H(r,z=0)=\frac{P}{2\pi w^2} e^{-\frac{r^2}{2 w^2}},\end{equation}
with absorbed power $P$ and standard deviation $w$. Here, $r$ is the radius from the center of the beam and $z$ points into the material. The steps presented to deduce the effective Raman temperature in this case are developed following Carslaw and Jaeger.\cite{Carslaw1986} We assume that the heat supplied by the Raman laser beam is only introduced in the plane $z=0$, which corresponds to a model where the excitation power is strongly absorbed at the surface. Within the material the temperature $T(\vec{r})$ must obey the stationary heat equation in cylindrical coordinates without heat sources
\begin{equation}\frac{\partial^2 T}{\partial r^2}+\frac{1}{r}\frac{\partial T}{\partial r}+	\frac{\partial^2 T}{\partial z^2}=0,\label{eq:RadHDE}\end{equation}
which is satisfied by 
\begin{equation}T \propto e^{-|\lambda| z} J_0(\lambda r)\end{equation}
for any $\lambda$ with $J_0(\lambda r)$ being the Bessel function of first kind and zeroth order. Circular heat flow in direction of the azimuthal angle $\phi$ can be neglected due to the symmetry of the problem. Equation~\ref{eq:RadHDE} is also satisfied by
\begin{equation}T = \int\limits_0^{\infty}  e^{-|\lambda| z} J_0(\lambda r) f(\lambda) \text{d} \lambda,\label{eq:T}\end{equation}
where $f(\lambda)$ is chosen to fulfil the boundary conditions. In our problem the Neumann boundary condition is given by the energy flow from the surface into the volume, introduced by the laser power density $H(r,z=0)$,
\begin{equation}-\kappa \left. \frac{\partial T}{\partial z}\right|_{z=0+} = \frac{P}{2\pi w^2} e^{-\frac{r^2}{2 w^2}}.\label{eq:bound}\end{equation}
Inserting equation~(\ref{eq:T}) into equation~(\ref{eq:bound}) leads to the condition 
\begin{equation}\kappa \int\limits_0^{\infty}  \lambda J_0(\lambda r) f(\lambda) \text{d} \lambda=\frac{P}{2\pi w^2} e^{-\frac{r^2}{2 w^2}} \end{equation}
for $f(\lambda)$. For the solution, the relation 
\begin{equation}\int\limits_0^{\infty}  x J_0(x r) e^{-\frac{w^2 x^2}{2}} \text{d} x= \frac{1}{w^2}e^{-\frac{r^2}{2w^2}}\label{eq:watson}\end{equation} 
is needed.\cite{Watson1995} Therefore, we can insert the function
\begin{equation} f(\lambda)= \frac{P}{2\pi \kappa} e^{-\frac{w^2\lambda^2}{2}}\label{eq:f}\end{equation}
into equation~(\ref{eq:T}), resulting in
\begin{equation}	T(r) = \frac{P}{2\pi \kappa} \int \limits_0^{\infty} J_0(\lambda r) e^{-\frac{w^2\lambda^2}{2}} \text{d} \lambda.	\label{eq:T2}\end{equation}
The effective Raman temperature $T_{\rm{Raman}}$ can then be obtained from equation~(\ref{eq:weighting}) and (\ref{eq:watson}) as
\begin{eqnarray}	T_{\rm{Raman}} &= T_{\rm{sink}}+\frac{1}{P} \int\limits_{\phi=0}^{2 \pi}  \int\limits_{r=0}^{\infty} T(r) H(r) \text{d} \phi r \text{d} r \nonumber \\
	&=T_{\rm{sink}}+ \frac{P}{4\sqrt{\pi} \kappa w}.	\label{eq:T3}\end{eqnarray}
For a homogeneous semi-infinite sample, excited by a gaussian shaped laser beam with strong absorption, the spatially constant thermal conductivity $\kappa$ is then given by
\begin{equation}	\kappa=\frac{P}{4\sqrt{\pi} \left( T_{\rm{Raman}}-T_{\rm{sink}}\right) w}.	\label{eq:kap}\end{equation}

To test the validity of equation~(\ref{eq:kap}) we investigate single-crystalline Si and Ge wafers. All Raman experiments in this work are performed using a Dilor spectrometer equipped with a \unit{1800}{l\per mm} grating and a liquid nitrogen cooled CCD detector. To map samples, an $x$-$y$ stage is used. An Ar ion laser operating at a wavelength of $\unit{514.5}{\nano\meter}$ excites Raman scattering. Various objectives are used for the micro Raman experiments, and their nearly gaussian spots were characterized by scanning the laser beam across the sharp edge of an evaporated Au film on top of a Si wafer, recording the decreasing Raman intensity of the Si LO mode. The spot width was obtained by deconvolution. For the  experiment on the wafers we use a $10\times$ objective with a spot standard deviation of $w=\unit{0.73}{\micro\meter}$. To enhance the accuracy we not only measure a single Raman spectrum for one excitation power, but perform series of measurements with different excitation powers. Then, (\ref{eq:kap}) changes to
\begin{equation}	\kappa=\frac{\frac{\partial \Delta k}{\partial T}}{4\sqrt{\pi} w\frac{\partial \Delta k}{\partial P}}.	\label{eq:kap2}\end{equation}
Due to the high thermal conductivity of the single-crystalline wafers, only a small temperature increase of less than \unit{60}{\kelvin} is observed during the experiment. Thus, we linearize the relation in figure~\ref{fig:SiLO} near room temperature and obtain $\frac{\partial \Delta k}{\partial T}=\unit{-0.0214}{cm^{-1}\per K}$ for Si. From the recorded power series on the single-crystalline Si wafer we obtain $\frac{\partial \Delta k}{\partial P}=\unit{-0.0245}{cm^{-1}\per mW}$, yielding $\kappa=\unit{168}{\watt\per\meter\usk\kelvin}$. With this result we only slightly overestimate literature values of $\kappa=\unit{156}{\watt\per\meter\usk\kelvin}$ and $\kappa=\unit{145}{\watt\per\meter\usk\kelvin}$, reported for Si around room temperature by references~\citenum{Glassbrenner1964} and \citenum{Maycock1967}, respectively. We have performed a similar experiment on a single-crystalline Ge wafer using $\frac{\partial \Delta k}{\partial P}=\unit{-0.0186}{cm^{-1}\per mW}$ from reference~\citenum{Menendez1984} and obtained $\kappa=\unit{49}{\watt\per\meter\usk\kelvin}$, in similarly good agreement with the value of $\unit{60}{\watt\per\meter\usk\kelvin}$ reported in reference \citenum{Glassbrenner1964} 

These results show that by using the Raman shift method applying equation~(\ref{eq:kap2}) one can measure the thermal conductivity of a homogeneous and 3-dimensional material with an accuracy of at least 10\%. The assumption of surface-near absorption of the excitation light, made during the deduction of equation~(\ref{eq:kap}), is fulfilled better for Ge, where the absorption coefficient for light at $\unit{514.5}{\nano\meter}$ is around $\alpha_{\rm{Ge}}=\unit{63\times10^4}{cm^{-1}}$,\cite{Humlicek1989} compared to Si with literature values around $\alpha_{\rm{Si}}=\unit{2\times10^4}{cm^{-1}}$.\cite{Humlicek1989,Sik1998,Aspnes1983} For a penetration depth in the range of or larger than the excitation laser beam, the effective area through which the heat is introduced into the material is enhanced, so that the thermal conductivity is over-estimated when equation~(\ref{eq:kap}) is applied. This may explain the tendency for our experiments on Si versus Ge wafers.
%
%
%
%
%
%
\subsection{In-Plane Conduction of Heat}\label{sec:2D}
\begin{figure}[tb]\centering\includegraphics{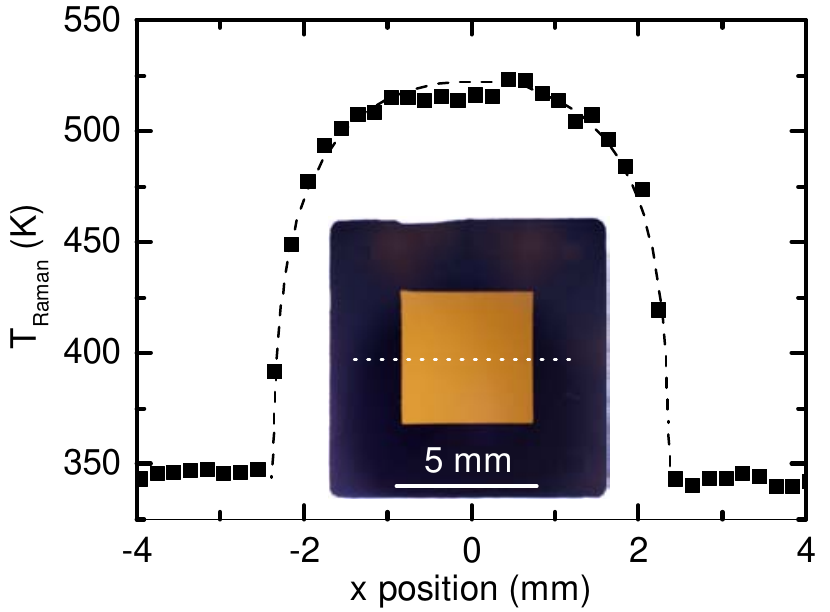}\caption{Raman temperature scan across a thin crystalline Si membrane of \unit{1.9}{\micro\meter} thickness. The full symbols show experimentally determined Raman temperatures as the excitation laser beam is scanned along the dashed line in the inset. $T_{\rm{Raman}}$ is increased on the suspended part and only weekly depends on the exact position in the center region. The dashed line is the result of a simulation with $\kappa_{\unit{300}{K}}=\unit{122}{\watt\per\meter\usk\kelvin}$, decreasing with temperature according to a power law with an exponent of $-1.15$\cite{Asheghi1997}. In the inset the bright region of the transmission optical microscopy image is the freely suspended part of the membrane, whereas in the dark region the membrane is supported by a $\unit{0.5}{mm}$ thick Si substrate.}\label{fig:2micrometerSi}\end{figure}

In the previous example the sample investigated was uniform so that Raman spectra taken at different positions on the bulk sample yield identical Raman temperatures. In this section we discuss an example where $T_{\rm{Raman}}$ depends on the position where it is measured due to the fact that although the thermal conductivity $\kappa$ can be expected to be homogeneous, the conductance is not. The sample is a \unit{1.9}{\micro\meter} thin and $10\times\unit{10}{mm^2}$ wide membrane of single-crystalline Si, which is carried by a \unit{0.5}{mm} thick Si support at the border. An optical micrograph of the sample in transmission is shown in the inset in figure~\ref{fig:2micrometerSi}. The membrane is freely suspended on an area of $4.8\times\unit{4.8}{mm^2}$. Using a $10\times$ objective resulting in a spot with a standard deviation of \unit{2.4}{\micro\meter} and a laser power of \unit{60}{\milli\watt}, the solid symbols in figure~\ref{fig:2micrometerSi} show an experimental Raman temperature scan across the sample, which was measured in vacuum. As soon as the excitation spot is on the freely suspended part of the membrane $T_{\rm{Raman}}$ increases. The heat absorbed in the membrane has to flow in-plane, which increases $T_{\rm{Raman}}$ when the excitation spot is moved away from the underlying support acting as the heat sink. In the center region of the membrane $T_{\rm{Raman}}$ is rather independent of the exact position. The variation of the experimental data in figure~\ref{fig:2micrometerSi} corresponds to an uncertainty of the determination of the thermal conductivity of the order of $10\%$. 

In contrast to the 3-dimensional heat flow problem in section~\ref{sec:3D}, here only 2-dimensional transport in the plane of the thin membrane is taken into account. Although the absorption follows an exponential dependence, the fact that the thickness of the film is of the order of $\alpha_{\rm{Si}}^{-1}$ allows to assume a homogeneous heating independent of the depth in the membrane, so that in our simulation no heat transport perpendicular to the membrane has to be considered. The high ratio of beam diameter and lateral size of the suspended membrane necessitates a large number of grid points in the simulation to correctly cover the temperature distribution at the excitation spot. Assuming a reflectivity of $38\%$,\cite{Humlicek1989,Sik1998,Aspnes1983} neglecting the temperature dependence of $\kappa$ would yield $\kappa=\unit{88}{\watt\per\meter\usk\kelvin}$ (not shown). However, in the temperature range relevant for this measurement and in the regime of thin films with a thickness of the order of micrometer, the thermal conductivity should be modelled by a power law dependence on temperature.\cite{Asheghi1997} The dashed line in figure~\ref{fig:2micrometerSi} is the result of our simulation of the Raman temperature across the suspended membrane with a room temperature thermal conductivity of $\kappa_{\unit{300}{K}}=\unit{122}{\watt\per\meter\usk\kelvin}$ and an exponent of approximately $-1.15$.\cite{Asheghi1997}

In thin films, phonon confinement effects decrease the thermal in-plane conductivity.\cite{Aksamija2010,Turney2010,McGaughey2011,Tang2011,Maznev2013} For films of the thickness investigated in our work, confinement is expected to reduce the conductivity at room temperature by a few percent.\cite{Asheghi1997,Asheghi1998,Ju1999,Liu2005,Maldovan2011,Chavez-Angel2014} This is in good agreement with the value of $\kappa$ obtained by our combination of Raman spectroscopy and simulation. In comparison to the bulk value obtained in the previous section this suggests that the Raman shift technique is applicable to thin films similarly well.
%
%
%
%
%
%
%
\section{Application of the Raman Shift Method}\label{sec:ApplicationofRSM}
In this section we will present applications of the Raman shift method to two Si-based samples, a bulk-like 3-dimensional material of dense nanocrystalline Si with considerable oxygen content and of a porous thin film of $\rm{Si}_{\rm{78}}\rm{Ge}_{\rm{22}}$.\cite{Stoib2012,Stoib2013,Stoib2014,Stein2011,Petermann2011,Kessler2013,Schierning2014} Both material systems are fabricated from the same type of raw material, which is a powder of gas phase synthesized nanocrystals of Si and SiGe, respectively. Among other possible applications, these materials are of considerable interest within the framework of thermoelectrics.\cite{Petermann2011,Stoib2012,Schierning2014} 

First, we show how to obtain $\kappa$ for a thin film which is suspended over a trench. Here, we assume a spatially constant thermal conductivity and set appropriate boundary conditions at the trench edges in the simulation. In such a case, the Raman temperature depends on the distance of the excitation spot to the heat sink as already seen in section~\ref{sec:2D}. Then, we investigate a bulk-nanocrystalline Si sample with a morphology suggesting a spatially varying $\kappa$ and analyse this with the model of a semi-infinite homogeneous material treated in section~\ref{sec:3D}. We take equation~(\ref{eq:kap}) as a basis for the evaluation and attribute different Raman temperatures to different local values of $\kappa$.
%
%
%
%
%
%
\subsection{Thin Laser-Sintered Nanoparticle Films }\label{sec:LaserSinteredFilms}
\begin{figure}[tb]\centering \includegraphics{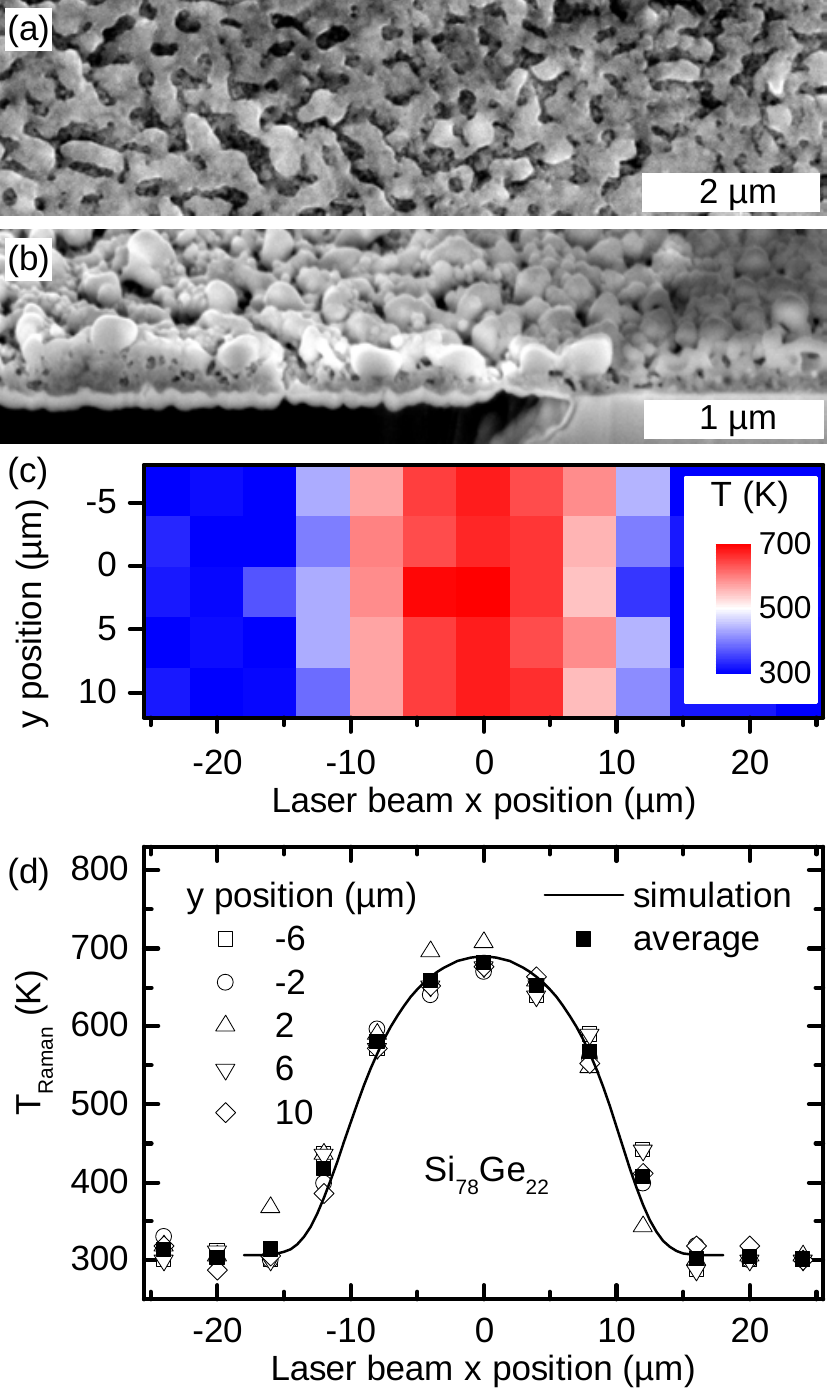}	\caption{(a)~Top view scanning electron microscopy (SEM) image of a thin film of laser-sintered $\rm{Si}_{\rm{78}}\rm{Ge}_{\rm{22}}$ nanoparticles. (b)~Side view SEM image of a such a film suspended on a trench. At the lower right corner, the underlying Ge wafer can be seen, acting as a heat sink. (c)~Colour coded Raman temperature map of the film on the trench. (d)~Raman temperature scans across the trench, together with the simulation of the Raman temperatures, shown as a solid line. Panel~(b) and (d) are reproduced with permission from \href{http://dx.doi.org/10.1063/1.4873539}{Appl. Phys. Lett. \textbf{104}, 161907}. Copyright 2014, AIP Publishing LLC.}	\label{fig:SiGebeides}\end{figure}

The thin film sample is a thin mesoporous film of $\rm{Si}_{\rm{78}}\rm{Ge}_{\rm{22}}$. It is fabricated by spin-coating a dispersion of \unit{23}{nm} diameter SiGe alloy nanoparticles to obtain films of \unit{300}{nm} thickness. The particles are heavily doped with 2\% P during their microwave plasma gas synthesis.\cite{Knipping2004,Stein2011,Petermann2011} This high doping level is typical for Si-based thermoelectric materials to optimise the power factor.\cite{Slack1991,Dismukes1964,Snyder2008,Schierning2014} After removal of the native oxide by hydrofluoric acid, the film is sintered in vacuum by a \unit{10}{ns} pulsed Nd:YAG laser operating at $\unit{532}{nm}$ with a fluence of \unit{100}{mJ\per cm^2}. The resulting mesoporous morphology is shown in a scanning electron microscopy (SEM) image in figure~\ref{fig:SiGebeides}(a). Further information on the fabrication and (thermo-)electric properties of such films can be found in references \citenum{Stoib2012} and \citenum{Stoib2013}.

In-plane measurements of $\kappa$ of as-fabricated thin films are hampered by the significant contribution of the substrate to the thermal conductance. Therefore, the film is transferred onto a support structure. This is a single-crystalline Ge wafer, into which trenches have been etched by reactive ion etching. Germanium was chosen as a heat sink material because its Raman spectrum does not overlap with the Si-Si phonon mode to be investigated. After the transfer, a focused laser beam was scanned along the border of the trench with a high fluence. This compacted the film and ensured a firmer attachment to the heat sink. Figure~\ref{fig:SiGebeides}(b) shows a detail of an SEM side view of such a suspended laser-sintered nanoparticle film.

In the Raman shift experiment of this sample we use an absorbed power of $P_{\rm{absorbed}}=\unit{12}{\micro\watt}$ for excitation and a $20\times$ objective with a spot standard deviation of \unit{0.64}{\micro\meter}. Due to the high surface area of the porous film the measurements are carried out in a vacuum chamber of a pressure of $p=\unit{10^{-1}}{mbar}$, which was found to be necessary to rule out spurious thermal conductance by contact to the surrounding ambient gas atmosphere. Because of the high Si content in the alloy the phonon mode that was used to extract $T_{\rm{Raman}}$ was the Si-Si vibration. Its temperature dependence in such SiGe alloys is very similar to that in pure Si and can be linearized in the region from room temperature to \unit{700}{K} by $\frac{\partial \Delta k}{\partial T}=\unit{-0.0229}{cm^{-1}\per K}$.\cite{Burke1993}

A Raman temperature map of the film with the suspended part in the center and the trench in $y$ direction is shown in figure~\ref{fig:SiGebeides}(c). In the suspended part of the film, the Raman temperature is as high as \unit{700}{K}, whereas the part of the film which is in direct contact with the supporting Ge wafer can efficiently conduct the heat introduced to the underlying heat sink, so that the Raman temperature stays close to room temperature. In figure~\ref{fig:SiGebeides}(d), the same data are shown in a more quantitative way. For different scans across the trench, the central Raman temperature varies by approximately 10\%, which gives the same estimate for the accuracy of determining $\kappa$ by the Raman shift method as discussed in section~\ref{sec:ModelSystems}. We also plot the mean value of all scans as filled squares, which is quite symmetric with respect to the center axis of the trench. 

In contrast to the measurement shown in figure~\ref{fig:2micrometerSi}, where the suspended part was \unit{4.8}{mm} wide in $x$ and $y$ direction, the suspended part of the laser-sintered thin film has a width of only \unit{19.5}{\micro\meter} (trench width) and a length of more than \unit{100}{\micro\meter}. Thus, the simulation grid used had an aspect ratio of 3:1, which was found to be of sufficient accuracy, neglecting the small fraction of heat transport in $y$ direction. 

The complete simulated profile of Raman temperatures across the trench assuming a temperature independent value of $\kappa$ is shown as solid line in figure~\ref{fig:SiGebeides}(d) and describes well the mean value of the experimental data within their experimental variation. In our simulation the porous film is treated as an effective medium and the solid line corresponds to an effective in-plane thermal conductivity of $\kappa_{\rm{eff}}=\unit{0.05}{\watt\per\meter\usk\kelvin}$ and a negligible contact resistance. In small grained and doped SiGe alloy thin films, it is justified to neglect the temperature dependence of the thermal conductivity,\cite{Stein2011,Steigmeier1964,McConnell2001} so that despite the large temperature differences in our experiment we obtain useful values for $\kappa_{\rm{eff}}$. Normalizing $\kappa_{\rm{eff}}$ by a factor $(1-\rm{porosity})$ with a typical porosity of 50\% for these laser-sintered thin films, we obtain the in-plane thermal conductivity usually given in literature.\cite{Boor2011,Tang2010} In the present case, this yields $\kappa_{\rm{normalized}}=\unit{0.1}{\watt\per\meter\usk\kelvin}$. Estimating all uncertainties entering the simulation we obtain a maximum thermal conductivity of $\kappa_{\rm{normalized}}^{\rm{max}}=\unit{0.3}{\watt\per\meter\usk\kelvin}$. 

It is generally believed that the mean free path of phonons is drastically reduced in materials with a hierarchy of scattering centers. The present sample exhibits such a disorder of different length scales, starting at the atomic scale due to alloy scattering in the SiGe alloy. Also some nanocrystals with a diameter of the order of \unit{25}{nm} survived the sintering process and are incorporated in the matrix. Typical for this type of laser-sintered material, grain boundaries between the grains of typically \unit{150}{nm} constitute larger scattering centers. For long wavelength phonons the mesoporous structure with typical structure sizes of \unit{300}{nm} is the relevant scatterer. In the laser-sintered mesoporous thin films a reduction of $\kappa$ by a factor of approximately 10-20 is observed, compared to nanograined but dense SiGe materials.\cite{Wang2008,Stein2011} Our value for $\kappa$ in mesoporous n-type doped SiGe is approximately a factor of 20 lower. Most likely percolation effects, which were intensively studied in pure Si materials and also affect electrical transport, are responsible for this additional reduction.\cite{Boor2011,Tang2010} 
%
%
%
%
%
%
\subsection{Bulk-Nanocrystalline Silicon}\label{sec:NCSi}
\begin{figure}[tb]	\centering	\includegraphics[width=0.3\textwidth]{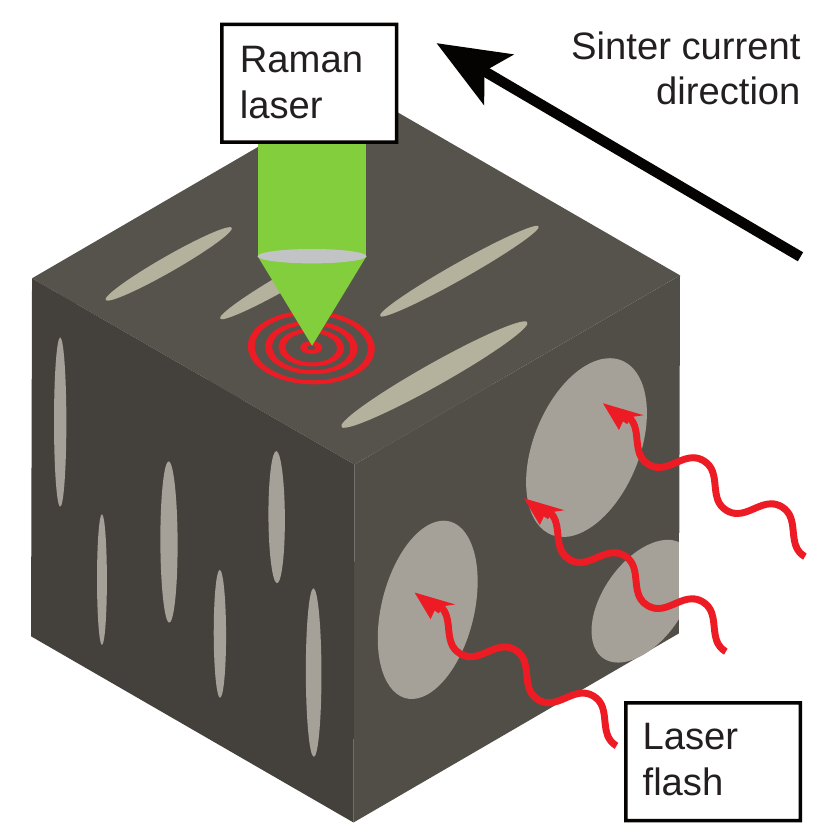}	\caption{Differences of the measurement geometries of the Raman shift method and the laser flash method, applied to bulk-nanocrystalline Si. The oxygen-rich areas of precipitates (grey) are disc shaped and lie perpendicularly to the laser flash measurement direction. For the microscopic Raman shift method, these precipitates play a less important role as barriers for thermal transport. The direction of the sinter current was the same as for the laser flash measurement.}\label{fig:SkizzeDiffRamanLaserFlash}\end{figure}

In this second application of the Raman shift method to Si-based materials, we investigate the local variation of $\kappa$ for bulk-nanocrystalline Si. The sample studied is synthesized from a powder of microwave plasma grown Si nanoparticles, which are doped with 1\% P in the gas phase and have a diameter of $22-\unit{25}{\nano\meter}$. The Si nanocrystal powder used for this sample was exposed on purpose to ambient oxygen for three weeks to obtain a significant oxygen content known to impact the microstructure. The powder was then pre-compacted and solidified by current-activated pressure-assisted densification, resulting in a slight increase in crystallite size to approximately \unit{50}{nm}.\cite{Petermann2011,Stein2011,Schierning2011} The direction of current in this sintering method leads to an anisotropy of the resulting material.\cite{Meseth2012} During densification, oxygen relocates within the nanoparticle network and forms mainly two types of oxygen-rich precipitates, small and rather spherical precipitates of approximately \unit{100}{nm} in size and larger agglomerates of such small precipitates forming larger structures of $\rm{SiO}_x$.\cite{Schierning2011,Meseth2012} The latter are shaped like a disc, with their axis pointing in the direction of the sinter current, and have diameters of several tens of microns and a thickness of approximately \unit{5}{\micro\meter}. The enriched oxygen content in the larger precipitates is accompanied by an enhanced porosity in this region.\cite{Schierning2011} Both, the different elemental composition and the different microstructure of the precipitates compared to the surrounding matrix, suggest a non-uniform thermal conductivity of the material. After densification, the sample investigated here was cut and polished by ion milling, so that the surface was flat on a tens of nanometer scale. Figure~\ref{fig:SkizzeDiffRamanLaserFlash} illustrates the orientation of the precipitates within the sample investigated. The Raman experiments were carried out on the polished top surface. Additional laser flash measurements of $\kappa$ were conducted from the orthogonal direction, due to geometrical restrictions of the sample. The direction of the sinter current was parallel to the direction of laser flash measurements. 

\begin{figure*}[t!]	\centering	\includegraphics{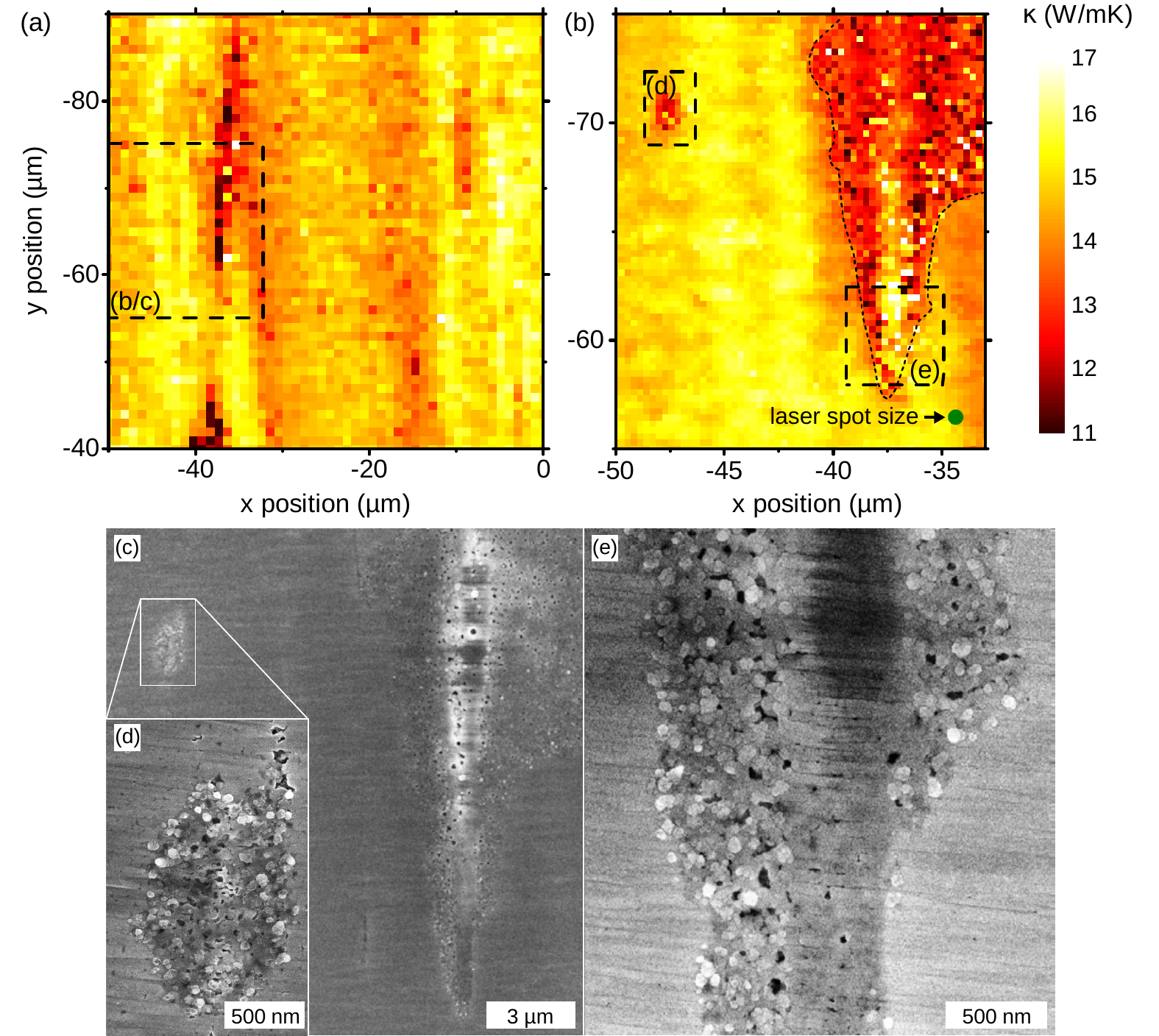}	\caption{Locally varying thermal conductivity of heavily P-doped bulk-nanocrystalline Si, obtained from a Raman temperature map and using equation~(\ref{eq:kap}). In panel~(a) an overview of the thermal conductivity map is shown. A zoom-in of in the dashed region in panel~(a) is shown in panel~(b), demonstrating a nearly micron resolution of the thermal conductivity mapping. The green dot in the lower right corner shows the gaussian full width of the laser beam used. The darker regions with a lower thermal conductivity coincide with oxygen-rich precipitates. Panel~(c) shows an SEM image of the area investigated in panel~(b). Panel~(d) and (e) show SEM details of precipitates, which are also marked by dashed lines in panel~(b). Outside the border of the precipitate in panel~(e), marked with a dotted line in panel~(b), the sample surface is flat.}\label{fig:Duisburg}\end{figure*}

The investigation of local variations of the thermal conductivity of this sample is based on the following procedure: Applying the Raman shift method, we first extract a Raman temperature map. Using an incident laser power of \unit{600}{mW} the sample is partly heated up to \unit{800}{\degree C}, so that we use $\frac{\partial \Delta k}{\partial T}=\unit{-0.0255}{cm^{-1}\per K}$ as a linear interpolation in figure~\ref{fig:SiLO}. The high signal-to-noise ratio in the Raman experiments allows to include the contribution of free charge carriers, introduced by the high amount of P and the strong illumination, in the evaluation of the Raman spectrum. Therefore, in contrast to the other experiments of this work, the Raman temperature is not determined experimentally from the maximum of the Raman line, but rather from a fit of a Fano lineshape to the spectra.\cite{Cerdeira1972,Chandrasekhar1978} Although the material is not homogeneous, we assume it to be homogeneous in the near field of the excitation laser beam. In these study we use a $100\times$ objective resulting in a gaussian beam of \unit{0.29}{\micro\meter} standard deviation, which is much smaller than average distances of the oxygen-rich precipitates. We again assume a reflectivity of $38\%$.\cite{Humlicek1989,Sik1998,Aspnes1983} By using equation~(\ref{eq:kap}) we then calculate a map of local thermal conductivities. 

Figure~\ref{fig:Duisburg} shows such a map of the thermal conductivity, using a different colour scale than in the maps of $T_{\rm{Raman}}$ discussed earlier, and the corresponding microstructure of the bulk-nanocrystalline Si sample as observed by SEM. In panel~(a) an overview map of $\kappa$ is shown. The map exhibits anisotropic structures which are elongated in $y$ direction and have a lower thermal conductivity compared to the surrounding Si matrix. The dashed rectangle in (a) is shown in panel~(b) with a higher resolution compared to panel~(a). Structures on the length scale of a micrometer can be discerned, which demonstrates that the measurement is capable to detect local variations in $\kappa$ close to the resolution limit given by the spot size. As a guide to the eye, the green dot in the lower right corner illustrates the full gaussian width of the laser excitation spot. 

The thermal conductivities obtained by the Raman shift experiment are in the range between 11 and \unit{17}{\watt\per\meter\usk\kelvin}. This is an order of magnitude lower compared to the values reported for undoped single-crystalline Si.\cite{Glassbrenner1964,Maycock1967} The extremely high content of P and the small grained nanostructure on a scale of \unit{50}{nm} resulting from sintering the small nanoparticles can be made responsible for this reduced thermal conductivity.\cite{Petermann2011,Schierning2014,Stein2011,Schwesig2011} The thermal conductivity of the very sample investigated here has also been characterized as a function of temperature using the laser flash method. At room temperature the laser flash method yields a thermal conductivity of $\kappa=\unit{9.5}{\watt\per\meter\usk\kelvin}$, which decreases to $\kappa=\unit{6.5}{\watt\per\meter\usk\kelvin}$ at \unit{1240}{K}. Thus, the temperature dependence is not pronounced, justifying the neglect of a temperature dependence of $\kappa$ when deducing equation~(\ref{eq:kap}) also for this type of sample. However, the values obtained for $\kappa$ obtained by the laser flash method are roughly a factor of 2 lower, compared to the results obtained by the Raman shift method. The most likely reason for this difference is the measurement geometry. As sketched in figure~\ref{fig:SkizzeDiffRamanLaserFlash}, for the laser flash measurements the heat flow was perpendicular to the disc shaped precipitates, making them a maximum barrier for heat transport. In the Raman shift method, the heat is spread radially into the material, with heat transport suffering only little from the alignment of the precipitates. Further reasons for the slightly different result are spurious thermal conduction by air during the measurements and the finite absorption coefficient of Si at the wavelength used, which leads to a slight over-estimation of $\kappa$ using equation~(\ref{eq:kap}) as discussed before.

To attribute the local variations in $\kappa$ observed in this material to structural features, we show SEM micrographs of the areas investigated by the Raman shift method in panel~(c) to (e) of figure~\ref{fig:Duisburg}. Panel~(c) shows the region investigated in panel~(b). The large structure on the right half of the panel can clearly be recovered in the SEM image. Also the smaller feature in the upper left corner of panel~(b) can be found in panel~(c), and is magnified in panel~(d). In contrast to the surrounding area, the surface of this feature is less flat and shows a porous interior. The same conclusion can be drawn from panel~(e), which shows the second rectangular area marked with dashed lines in panel~(b). A similar porosity as in the small feature can be found here. Energy dispersive X-ray scans across the structure in panel~(e) confirm that the oxygen content in the porous region is enhanced by at least a factor of 4.\cite{Meseth2012} Correlating the SEM image in panel~(e) to the thermal conductivity map in panel~(b) suggests that the porous regions clearly visible in SEM exhibit a lower thermal conductivity compared to the surrounding area. At least in principle, this apparently lower thermal conductivity could arise from the local increase of the absorbed laser power, which in turn could be caused by the roughness of the surface visible in the SEM micrographs.\cite{Algasinger2013} However, since strong variations in $\kappa$ are also found for flat parts of the bulk-nanocrystalline sample studied, it can be concluded that the contrast in the maps of thermal conductivity originates to a significant part from the locally varying thermal conductivity. 
%
%
%
%
%
%
\section{Summary and Conclusion}\label{sec:Conclusion}
We showed that by performing a micro Raman scattering experiment where the laser simultaneously acts as a thermal excitation source and as a thermometer, using the temperature dependence of the energy of Raman active phonon modes, one can determine the thermal conductance of a specimen. Knowing or simulating the geometry of heat propagation from the excitation spot to the heat sink is key to obtaining reliable data on the thermal conductivity. We discussed that it is necessary to take the non-homogeneous temperature distribution beneath the excitation spot into account to correctly interpret the effective temperature deduced from the Raman spectrum. Applying this Raman shift method to both 3-dimensional heat flow into a semi-infinite homogeneous material and to 2-dimensional heat transport in a suspended thin film we experimentally validated the technique and its analysis. Finally, we used the Raman shift method to determine the thermal in-plane conductivity for laser-sintered thin films of $\rm{Si}_{\rm{78}}\rm{Ge}_{\rm{22}}$ nanoparticles. Assuming a spatially constant $\kappa$ and attributing an increased temperature solely to the locally varying distance to the heat sink, we demonstrated that the Raman shift method can measure porosity-normalized values of the thermal conductivity as low as $\unit{0.1}{\watt\per\meter\usk\kelvin}$. As a second application, we investigated local variations of the thermal conductivity of a 3-dimensional bulk-nanocrystalline Si sample exhibiting microscopic $\rm{SiO}_x$ precipitates in SEM investigations. Here, the Raman shift method is able to measure local variations of the thermal conductivity by more than 40\% between oxygen-rich porous regions and dense regions with reduced oxygen content with a spatial resolution of the spot size of the exciting laser beam.

Obtaining reliable quantitative information on the local thermal conductivity requires sound knowledge on three major parameters, which are the intensity profile of the exciting laser beam, the geometry of thermal transport and the absorbed optical heating power. The latter turns out to be the most critical parameter for samples with complex microstructure and can be challenging to determine. Rough surfaces or porous materials, often accompanied by a spatial variation of the elemental composition, can require to base the evaluation of the results of the Raman shift method on assumptions on absorption coefficients and reflectivities, since the direct measurement of reflected and transmitted excitation power is difficult in many sample geometries. The vector field of heat propagation can only be calculated in very rare cases analytically. Therefore, numerical simulations need to be performed to solve the heat diffusion equation which involves considerable computation effort, especially when the problem cannot be reduced to two dimensions. Finally, although the intensity profile of the laser beam used can easily be accessed experimentally, the implications of a spatially inhomogeneous excitation combined with a non-homogeneous temperature distribution on the resulting Raman spectrum measured can be manifold. This includes Raman scattering cross sections, line shapes, absorption profiles or collection efficiencies. However, a set of reasonable assumptions can make the Raman shift method a straight forward method.

This study demonstrated that the variety of materials systems and sample geometries that can be investigated by the Raman shift method without mechanical contact makes the method a versatile and powerful tool to obtain thermal information of small scale and complex materials systems. With that, the method complements well more traditional and established tools and enables insight into thermal transport on a micrometer scale.
%
%
%
%
%
%
\section*{Acknowledgments}
We acknowledge funding by the German Research Foundation DFG via the priority program SPP~1386 "Nanostructured Thermoelectrics" and additional support by the Bavarian State Ministry of the Environment and Consumer Protection via the project "Umwelt Nanotech".
\end{document}